\documentclass[11pt]{cernrep}
\usepackage[dvips]{graphicx}

\def\x{{\bf x}}

\def\k{{\bf k}}

\def\o{\omega}
\def\llangle{\left\langle}
\def\rrangle{\right\rangle}

\def\Im{{\rm Im}}

\def\st{\begin{equation}}
\def\stp{\end{equation}}
\def\bg{\begin{eqnarray}}
\def\nd{\end{eqnarray}}
\def\JJ{{\scriptscriptstyle JJ}}
\def\NN{{\scriptscriptstyle NN}}

\begin{document}

\title{Heavy Quark Diffusion and Lattice Correlators}

\author{ P.~Petreczky$^{1,2}$, K. Petrov$^2$, D. Teaney$^3$ and A.~Velytsky$^{4}$}
\institute{
$^1$RIKEN-BNL Research Center, Brookhaven National Laboratory, Upton, NY, 11973,  USA\\
$^2$Physics Department, Brookhaven National Laboratory, Upton, NY, 11973,  USA\\
$^3$Department of Physics and Astronomy, SUNY at Stony Brook, Stony Brook, New York 11764, USA\\
$^4$Department of Physics and Astronomy, UCLA, Los Angeles, CA 90095-1547, USA
}

\maketitle

\begin{abstract}
We study charmonia correlators at finite temperature.
We analyze to what extent heavy quarkonia correlators
are sensitive to the effect of heavy quark transport and 
whether it is possible to constrain the heavy quark diffusion constant
by lattice calculations. Preliminary lattice calculations of quarkonia 
correlators performed on anisotropic lattices show that they are sensitive
to the effect of heavy quark transport, but much detailed calculations
are required to constrain the value of the heavy quark diffusion constant.
\end{abstract}

\section{Introduction}
There are plenty of experimental evidence that strongly interacting matter
at high energy density has been produced at RHIC \cite{Bellwied:2005kq,Adcox:2004mh}.
One of the most exciting results
from  RHIC  so far is the large azimuthal anisotropy of light hadrons
with respect to the reaction plane, known as elliptic flow. The observed 
elliptic flow is well described by ideal hydrodynamics 
\cite{Hirano:2004er,Teaney:2001av,Kolb:2000fh} suggesting early equilibration 
of the produced matter and very short transport mean free path.
This interpretation of the experimental data can be challenged by measuring elliptic
flow of charm and bottom mesons \cite{Laue:2004tf,Adler:2005ab,Adler:2004ta}. 
The first experimental results show a non-zero elliptic
flow for these heavy mesons.
Naively, since the
quark mass is significantly larger than the temperature of the medium,
the mean free pass  of heavy mesons is $\sim M/T$ longer
than the light hadron mean free path. Quantitatively the mean free path
is described by the heavy quark diffusion constant which can be defined
through the diffusion equation for the heavy quark number density $N(\x,t)$,
${\partial}_t N + D \nabla^2 N = 0$.
If the heavy quark diffusion constant $D \ge 1/T$, the predicted heavy quark elliptic
flow will be too small and in contradiction with current experimental data 
\cite{Moore:2004tg}.
 
Kubo formulas relate hydrodynamic 
transport coefficients  to the small frequency behavior
of real time correlation functions \cite{Forster,BooneYip}.
Correlation functions in real
time are in turn related to correlation functions in imaginary time by analytic
continuation. Karsch and Wyld \cite{karschwyld} 
first attempted to use this connection
to extract the shear viscosity of QCD from the lattice.
More recently, additional attempts to extract the shear
viscosity \cite{nakamura97,nakamura05} and 
electric conductivity \cite{gupta03} have been made.
It turns out that Euclidean correlations functions 
are remarkably insensitive 
to transport coefficients. For weakly coupled field theories 
this has been discussed by Aarts and Martinez Resco \cite{aarts02}.
For this reason, only precise lattice data and a comprehensive
understanding of the different contributions to the Euclidean
correlator can constrain the transport coefficients.
It appears that heavy quarkonia correlators are the likely candidates
for meeting this conditions.

\section{Euclidean and real time correlators}

On the lattice we calculate correlation function of local meson
operators ( currents ) $J_E^h(\x,\tau)=\bar q(\x,\tau) \Gamma_h q(\x,\tau)$
at finite temperature
$$
   G^h(\k,\tau,T) = \int d^3\x\, e^{i\k \cdot \x} \,
  \llangle J_{E}^h(\x, \tau) J_{E}^h(0, 0) \rrangle,
$$
with $\Gamma_h$ being some combination of the  Dirac matrices. 
This correlation function is related to the real time correlation functions
$D_h^{>}(\x,t,T)=\langle J^h(\x,t) J^h(0,0) \rangle$,
$D_h^{<}(\x,t,T)=\langle J^h(0,0) J^h(\x,t) \rangle$.

The most important channels for our further discussion
are the pseudo-scalar, $\Gamma_h=\gamma_5$ and the vector, $\Gamma_h=\gamma_{\mu}$ 
channels. In the vector channel the Euclidean correlators are related to
density-density correlator $D^{>}_{NN}=\langle N(\x,t) N(0,0) \rangle$ and
current-current correlators $D^{>\, ij}_{JJ}=\langle J^i(\x,t) J^j(0,0) \rangle$,
\begin{equation}
G^{00}({\bf x},\tau,T)=\llangle J^0_{E}({\bf x},\tau) J^0_{E}(0,0) \rrangle=
-D_\NN^>({\bf x}, -i \tau,T) \, ,
\label{nncorr}
\end{equation}
\begin{equation}
G^{ij}({\bf x},\tau,T)=\llangle J^i_{E}({\bf x},\tau) J^{j}_E(0,0) \rrangle=
D_\JJ^{>,ij}(\x, -i \tau,T) \, .
\end{equation}
Similarly for the pseudo-scalar channel 
$G^5({\bf x},\tau,T)=\llangle J^5_{E}({\bf x},\tau) J^5_E(0,0) \rrangle=
D_5^{>}(\x, -i \tau,T) \,$ . The minus sign in Eq. (\ref{nncorr}) comes
from the relation $A^0=-iA^0_E$ between the temporal component of the
vector in Minkowski space and Euclidean space, in particular
$x^0=-ix^0_E=-i\tau$.
The spectral function is defined through Fourier transform of 
$D_h^{<}$ and $D_h^{>}$ or equivalently as imaginary mart of the retarded
correlator $\chi_h(\k,\omega)$
\st
   \rho_h(\k,\omega,T)  = \frac{D_h^{>}(\k, \omega,T) - D_h^{<}(\k, \omega,T) }{2\pi} 
=\frac{1}{\pi} \Im \chi_h(\k,\omega,T)\, .
\stp 
Using the Kubo-Martin Schwinger (KMS) relation 
$D_h^{>}(\k,t) = D_h^{<}(\k,t + i/T)$ 
one discovers the relation between the spectral density 
and the Euclidean correlator,
\st
  G^h(\k,\tau,T) = (-i)^{r} \int_0^{\infty} d\omega \, \rho^h(\k,\omega,T) 
\frac{ \cosh\left( \omega\left(\tau -\frac{1}{2T}\right)\right) }
{\sinh\left(\frac{\omega}{2T}\right) } \, .
\label{spectral_rep}
\stp
Here $r$ is number of zeros in the space-time indexes.

\section{Lattice results on the charmonia correlators and spectral functions}

Charmonia correlators have been studied in lattice QCD and the corresponding
spectral functions were reconstructed using the Maximal Entropy Method (MEM)
\cite{umeda02,asakawa04,datta04}.
These studies showed that the 1S states ($\eta_c$ and $J/\psi$) survive in
the plasma up to temperatures as high as $1.6T_c$. Though it is quite 
difficult to reliably reconstruct the spectral functions, the temperature
dependence of the correlators can be determined quite precisely \cite{datta04}. 

We calculated charmonia correlators on quenched anisotropic lattices using
the Fermilab formulation for heavy quarks \cite{chen01}. Calculation were
done at $\beta=6.5$ and $\xi=a_s/a_t=4$, corresponding to temporal lattice
spacing $a_t^{-1}=14.12$GeV when we set the spatial lattice spacing $a_s$ 
using the Sommer scale $r_0=0.5$fm. 
We collected about $1000$ gauge configurations at each temperature.
From Eq. (\ref{spectral_rep}) it is clear
that the temperature dependence of the correlator $G(\k,\tau,T)$ comes from  
temperature dependence of the spectral function and temperature dependence of
the kernel 
$K(\tau,\omega,T)=
\frac{ \cosh\left( \omega\left(\tau -\frac{1}{2T}\right)\right) }
{\sinh\left(\frac{\omega}{2T}\right) }$. To separate out the trivial 
temperature dependence due to the kernel $K(\tau,\omega,T)$ following
\cite{datta04} we introduce the reconstructed correlator
\st
G_{{\rm rec}}^h(\k,\tau,T)=\int_0^{\infty}d\omega\, \rho^h(\k,\omega,T=0) \,
K(\tau,\o,T).
\label{grec}
\stp
If the charmonia spectral function do not change across the deconfinement
transition temperature $T_c$ we expect $G^h/G^h_{\rm rec}\simeq 1$.
In Fig. 1 we show the temperature dependence of $G^h/G^h_{\rm rec}$ for
pseudo-scalar and vector channels at zero spatial 
momentum $\k=0$. In the vector channel we show both
sum over all spatial components $\sum_i G^{ii}$ and the sum over
all four components $\sum_{\mu} G^{\mu \mu}$. We see that the temperature
dependence of the vector and pseudo-scalar correlarors is quite different.
For $T=1.5T_c$ we see only very small deviations from unity for $G^h/G^h_{\rm rec}        $
in the pseudo-scalar channel while 
significant deviations are seen in the vector channel. 
In fact similar temperature dependence of the vector correlator was seen
in the  previous study based on fine isotropic lattices 
\cite{datta04,dattasewm04}.
This is quite
unexpected as $\eta_c$ and $J/\psi$ should have similar properties both
in the vacuum and in the medium. We will give an explanation for this difference
in the next section in terms of heavy quark transport.

\section{Spectral functions at low energies and heavy quark transport}

As vector current is a conserved current there should be transport
contribution  to the corresponding spectral function. 
In general the vector spectral function can be decomposed in terms
of transverse $\rho^T(\k,\o,T)$ and longitudinal $\rho_L(\k,\o,T)$ components. 
Since the heavy quark mass is much larger than the temperature
$M \gg T$ we can write
$\rho^{L,T}(\k,\omega,T)=\rho_{\rm low}^{L,T}(\k,\omega,T)+\rho_{\rm high}^{L,T}(\k,\omega,T)$,
where $\rho^{L,T}_{\rm high}(\k,\omega,T)$ contains the resonances and the 
continuum, and is non-zero for energies $\omega \sim 2M$, and
$\rho^{L,T}_{\rm low}(\k,\omega,T)$ is the transport contribution.
The simplest way to estimate 
$\rho^{L,T}_{\rm low}(\k,\omega,T)$ is to evaluate the
vector correlator at 1-loop level \cite{derek05}. 
In the $\k \rightarrow 0$ limit we have 
$\rho^T(0,\o,T)=\rho^L(0,\o,T)=\rho^{ii}(0,\o,T)$,  and 
considering small energies, $\omega \ll T$ we get
\st
\rho^{ii}_{\rm low}(0,\omega,T)=
\chi_s(T) \frac{T}{M} \omega \delta(\omega),~
\rho_{\rm low}^{00}(0,\omega,T)=\chi_s(T) \omega \delta(\omega).
\stp
Here $\chi_s(T)$ is the static charm number susceptibility, which in
the limit $M \gg T$ is given by
$\chi_s(T)=12 \left(\frac{MT}{2 \pi}\right)^{3/2} e^{-M/T}$.
Thus at finite temperature we expect that the $\sum_i G^{ii}$
should be enhanced by a constant contribution 
$3 \chi_s(T) T/M$ relative to its $T=0$ value, 
while the $\sum_{\mu} G^{\mu \mu}$ should reduced by
$-\chi_s(T) (1-3 T/M)$ (recall Eq. (\ref{nncorr})). 
This is exactly what the lattice data
in Fig. 1 show. Furthermore, from data on $\sum_i G^{ii}$ and
$\sum_{\mu} G^{\mu \mu}$ we can estimate that $M/T \simeq 6$ at 
$1.5T_c$. The 1-loop
result for the vector correlator can be also obtained using collision-less
Boltzmann equation describing free streaming of heavy quarks with no interaction with
the plasma \cite{derek05}. This 1-loop contribution happens to dominate the
transport part of the Euclidean correlator \cite{derek05}.
\begin{figure}
\includegraphics[width=2.9in]{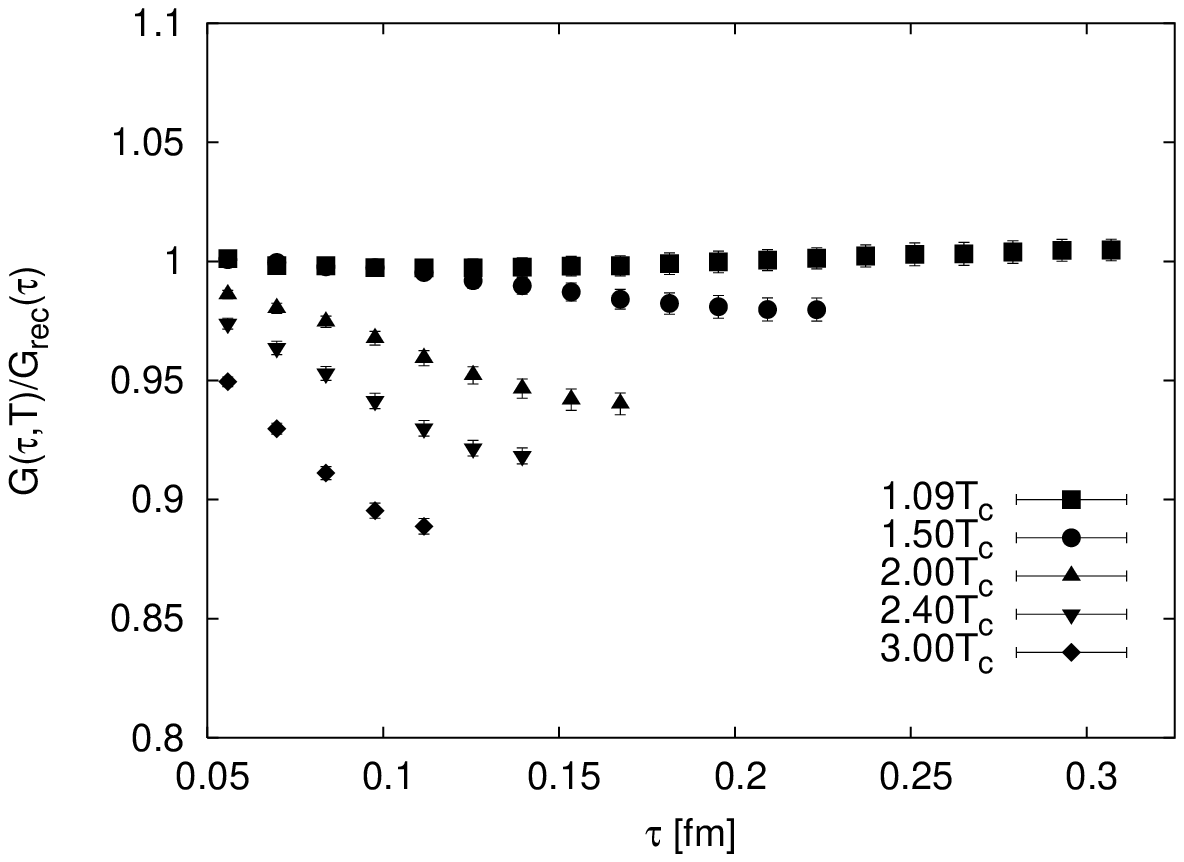}
\includegraphics[width=2.9in]{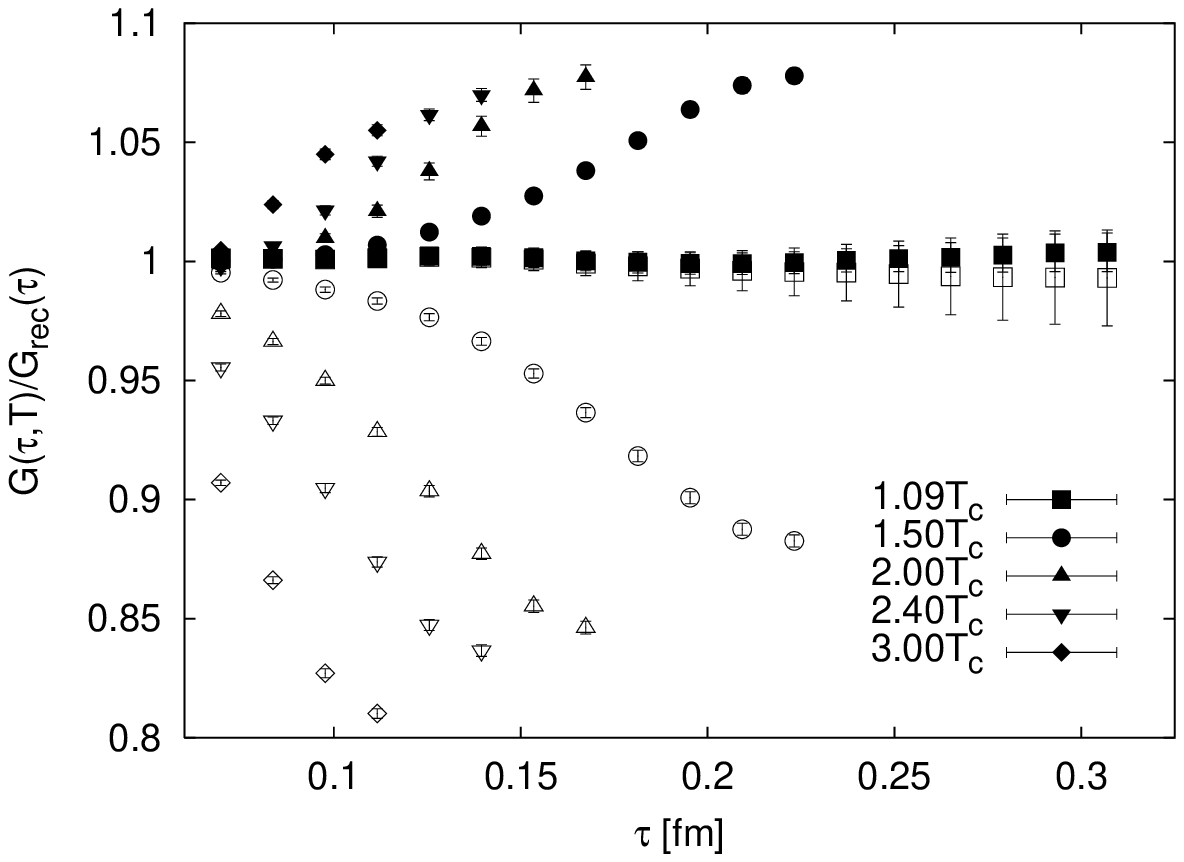}
\caption{The ratio $G/G_{\rm rec}$ for pseudo-scalar (left) and vector (right)
channels at $\k=0$. For the vector case we show both the sum over spatial components
(filled symbols) and all four components (open symbols). }
\end{figure}

\section{Effective Langevin equation for heavy quark transport}
We have seen that the leading transport contribution to the
Euclidean correlator is just a constant and corresponds to
free streaming of heavy quarks. To get the transport coefficient
we need to include the effect of heavy quark interactions with 
the medium. It is very difficult problem in general.
Luckily, the case of heavy quarks is special since  the time scale for
diffusion, $~M/T^2$, is much longer than any other time scale in the problem.
In terms of the spectral functions, this separation 
means that transport processes
contribute at small energy, $\omega \sim T^2/M$.
For this reason
we will assume that the Langevin equations provide a good 
macroscopic description of the dynamics of charm quarks \cite{Moore:2004tg},
\st
\label{newton_langevin}
\frac{dx^i}{dt} = \frac{p^i}{M}\,,  \frac{dp^i}{dt} = \xi^i(t) - \eta p^i\,, \qquad 
\langle \xi^i(t) \xi^j(t') \rangle = \kappa \delta^{ij} \delta(t-t') \, .
\stp
The drag and fluctuation coefficients are related by the 
fluctuation dissipation relation 
$\eta = \frac{\kappa}{2 MT}$.
For time scales which are much larger than $1/\eta$ the 
heavy quark number density obeys ordinary diffusion equation
$\partial_t N + D \nabla^2 N = 0$.
The drag coefficient $\eta$ can be related to the diffusion coefficient 
through the Einstein relation $D = \frac{T}{M \eta} = \frac{2 T^2}{\kappa}$.

\begin{figure}
\hspace*{-0.5cm}
\includegraphics[width=3.0in]{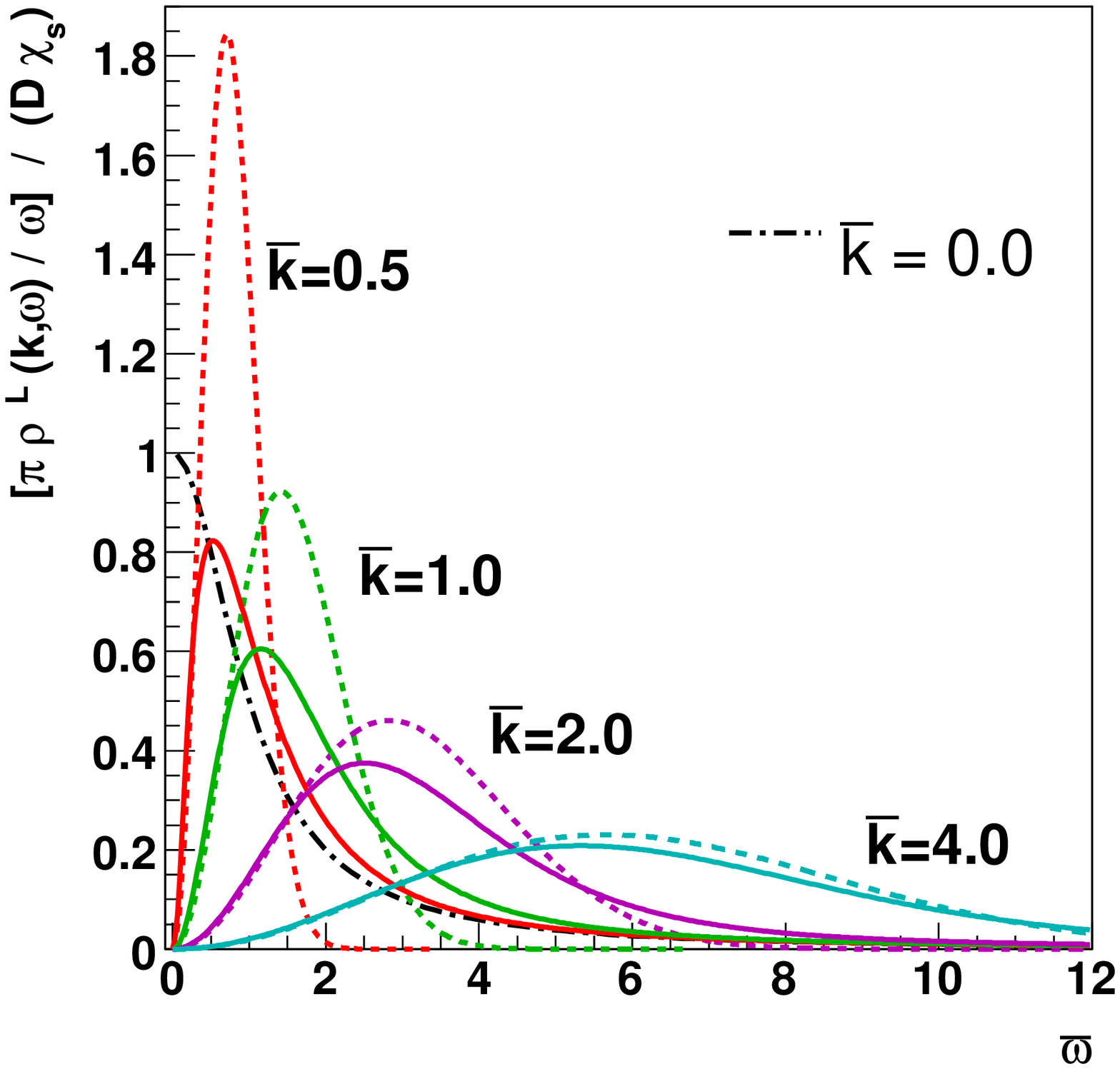}
\hspace*{-0.5cm}
\includegraphics[width=2.8in]{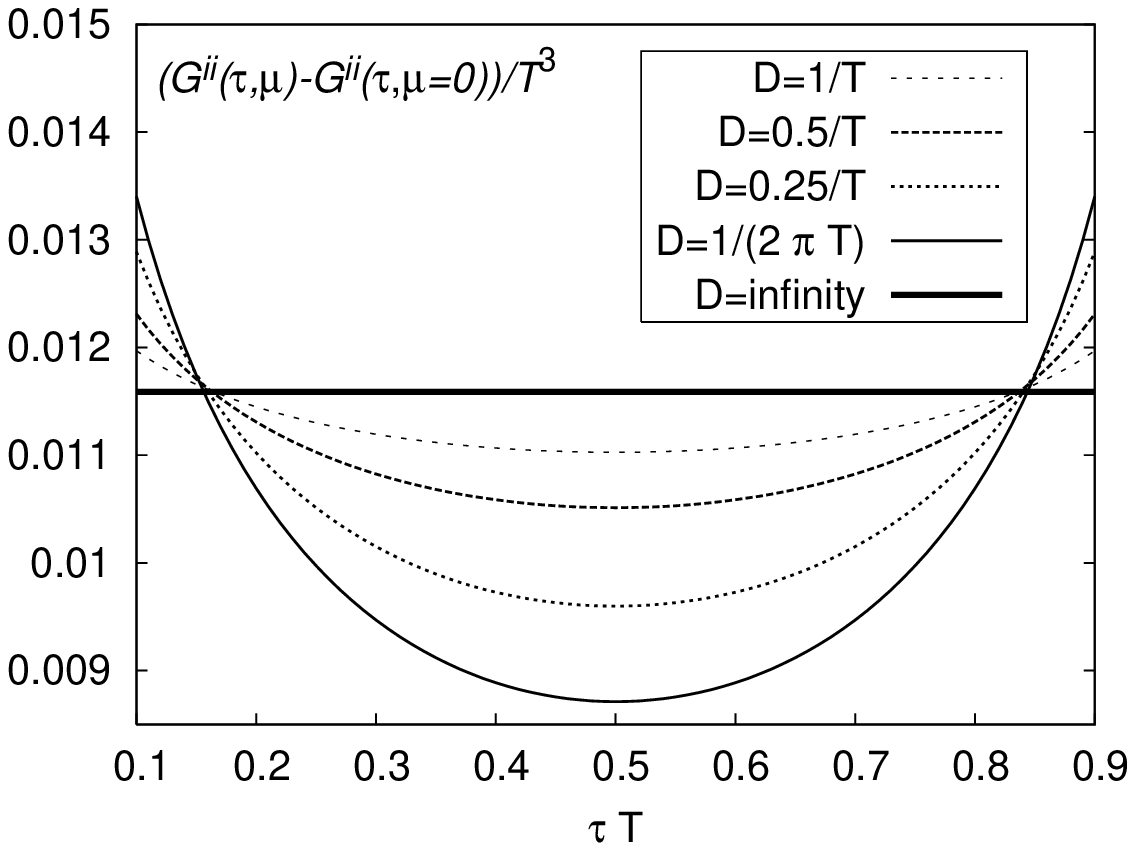}
\caption{The low energy contribution to the longitudinal 
spectral function $\rho^L_{\rm low}(\k,\o,T)$ (left) and
the low energy contribution to the correlator (right).
Here $\bar \o=\o D (M/T)$ and $\bar k= k D \sqrt{D/T}$.
}
\end{figure}
The effective Langevin theory can be derived from 
kinetic theory in the weak coupling limit \cite{Moore:2004tg}
and probably is
adequate for describing heavy quark diffusion even for
strongly interacting plasma.
The Langevin equations make a definite prediction for
the retarded correlator $\chi_h$ at small $\omega$ and thus 
for the transport part of the spectral functions \cite{derek05}.
The results of calculation for the longitudinal spectral functions is shown
in Fig. 2. For the case of zero spatial momentum $\k=0$ we 
have 
\st
    \frac{ \rho^{ii}_{\rm low}(0, \omega,T)}{\omega}=
          \chi_s(T)\,\frac{T}{M} 
\frac{1}{\pi}\, \frac{ \eta }{\omega^2 + \eta^2}\, ,
\frac{ \rho_{\rm low}^{00}(0, \omega,T)}{\omega}=\chi_s(T) \delta(\omega).
\label{spf_low}
\stp
From Eq. (\ref{spf_low}) it is clear that to calculate the
transport coefficient we have to determine the curvature of
$G^{ii}(\k=0,\tau,T)$ at $\tau=1/(2T)$ due to the low energy part of the
spectral function $\rho^{ii}_{\rm low}$.
If $\rho^{ii}_{\rm low}$ was the only contribution to the spectral function and
$\eta=0$ the correlator would be constant.
The question is how to determine the small curvature in 
$G^{ii}(\k=0,\tau,T)$, arising from finite value of $\eta$, from
the curvature arising from the resonance and continuum
contributions. This can be done 
by introducing a small chemical potential for the heavy quark, 
$\mu \ll M$. Since the transport
contribution is proportional to $\chi_s$, the small
chemical potential will enhance the transport by factor of $\cosh(\mu/T)$ \cite{derek05}.
The small charm chemical potential will not affect the resonance and 
continuum contributions to the spectral function 
to leading order in the heavy
quark density, $\sim e^{-(M - \mu)/T}$.
Thus we expect
that
\begin{eqnarray}
 \delta G^{ii} &\equiv& G^{ii}(\tau,T,\mu)-G^{ii}(\tau,T,0)\nonumber \\
  & \simeq &
(\cosh(\mu/T)-1) 
\int_0^{\infty} d \omega \left. \rho^{ii}_{\rm low}\right|_{\mu=0}(0,\omega,T)
\frac{\cosh(\omega(\tau-1/(2T)))}{\sinh(\omega/(2T))},
\end{eqnarray}
is largely insensitive to the high frequency behavior of
the spectral function. 
In Fig. 2 we show $\delta G^{ii}$ quantity for several values of $D$.
The numerical results on the vector correlator
show that it can be calculated with $0.5\%$ statistical accuracy. 
Thus if similar numerical accuracy can be achieved for the difference
$\delta G^{ii}$, the curvature and thus the $\eta$, or equivalently $D$ can be estimated
in lattice QCD.

\section*{Aknowledgements}
This work was supported by U.S. Department of Energy under 
contract DE-AC02-98CH1086 and through the SciDAC program.
D.T. was supported by the grant DE-FG02-88ER40388 and DE-FG03-97ER4014
of U.S. Department of Energy, A.V. was partly supported by NSF-PHY-0309362.

\end{document}